# Structural, Elastic, Electronic, Thermodynamic, and Optical Properties of Layered BaPd$_2$As$_2$ Pnictide Superconductor: a First Principles Investigation

F. Parvin and S.H. Naqib*

Department of Physics, University of Rajshahi, Rajshahi-6205, Bangladesh

*Corresponding author salehnaqib@yahoo.com

**Abstract**

BaPd$_2$As$_2$ belongs to 122 pnictide group, famous for exhibiting superconductivity at high temperatures and other electronic correlations. Unlike the 122 iron arsenides, BaPd$_2$As$_2$ exhibits superconductivity at low temperature when realized in theThCr$_2$Si$_2$ type structure (I4/mmm). We have performed density functional theory (DFT) based calculations to investigate the structural, elastic, electronic, thermodynamic, and optical properties of BaPd$_2$As$_2$ in this study. The structural, elastic, and the band structure features are compared with the available experimental and theoretical results. Pressure and temperature dependences of various important thermodynamic functions, e.g., bulk modulus, specific heats at constant pressure and volume, coefficient of volume thermal expansion, and Debye temperature are studied in details for the first time. The optical parameters of BaPd$_2$As$_2$ are also studied in details for the first time. The optical properties compliment the electronic band structure characteristics. Optical constants show significant dependence of the state of polarization of the incident electric field. BaPd$_2$As$_2$ exhibits high reflectance in the infrared and near-visible region and strongly absorbs the ultraviolet radiation. The relevance of the electronic energy density of states and the characteristic phonon frequency to superconductivity in BaPd$_2$As$_2$ is also discussed.

## 1 Introduction

Discovery of superconductivity in layered iron arsenides [1], the so-called 122 compounds with chemical formula AFe$_2$As$_2$ (A = Ca, Sr, Ba, etc.) and ThCr$_2$Si$_2$-type crystal structure, having high superconducting transition temperature have attracted enormous interest of the condensed matter physics community [2 – 8]. Beside high superconducting (SC) transition temperatures, these iron arsenides usually show spin density wave correlations coexisting and competing with superconductivity and semi-metallic behavior in the normal state [9]. The pairing mechanism leading to superconductivity in these 122 phases is a matter of intensive study.

It is known that many other transition metals crystallizes in the $ThCr_2Si_2$-type simple body centered tetragonal form [10]. Therefore, a search of superconductivity in iron free isostructural ternaries is a viable route for the condensed matter community. Indeed, in recent times a number of iron free superconductors have been discovered with $ThCr_2Si_2$-type layered structure [10 – 13].

The compound under study, $BaPd_2As_2$, is a Fe-free transition metal ternary. This compound is found to crystallize in three different forms: the $ThCr_2Si_2$-type structure (space group I4/mmm, No. 139), $CeMg_2Si_2$-type structure (space group P4/mmm, No. 123), and the intergrowth structure of the $CaBe_2Ge_2$-typeand CeMg2Si$_2$-type (I4/mmm) [11]. Superconductivity is seen in the $ThCr_2Si_2$-type structure which consists of $PdAs_4$tetrahedra, while the non superconducting$CeMg_2Si_2$-type structure consists of $PdAs_4$ planar squares. The third of the polymorphs, the intergrowth structure, consists of both $Pd_4As$ tetrahedra and $PdAs_4$ planar squares. Guo et al. [10] first reported superconductivity in $BaPd_2As$ with $T_c$ = 3.85 K for the $ThCr_2Si_2$-type structure. This superconducting transition temperature of $BaPd_2As_2$is significantly higher compared with those of 1.27 K and 0.92 K for isostructural and isoelectronic $CaPd_2As_2$ and $SrPd_2As_2$, respectively [12]. This has made $BaPd_2As_2$an interesting system to study [10 – 15]. Anand et al. [12] have studied the transport, magnetic and thermodynamical properties of superconducting $CaPd_2As_2$ and $SrPd_2As_2$ phases together with the non-SC polymorph of $BaPd_2As_2$, experimentally. Analysis of experimental parameters shows that $CaPd_2As_2$ and $SrPd_2As_2$ are weakly coupled electron-phonon superconductors with s-wave order parameter. Shein et al. [13] have investigated the elastic and electronic band structure properties of $CaPd_2As_2$ and $SrPd_2As_2$ superconductors and the non-SC $BaPd_2As_2$ with $CeMg_2Si_2$-type structure. Super conductivity in $BaPd_2As_2$ was reported first by Guo et al. [10]. Chen [14] carried out lattice dynamical calculations to study the electron-phonon coupling and its relevance to superconductivity in $BaPd_2As_2$ at relatively higher temperature. Kudo et al. [11] found a high electron phonon coupling constant ($\lambda_{e-ph}$), dominated by soft phonon modes in this SC compound. Very recently Abdel-Hafiez et al. [15] studied the physical properties of $BaPd_2As_2$ under pressure both experimentally and theoretically. This study indicates that even though$BaPd_2As_2$ is a layered material with quasi-2D features, superconductivity is isotropic in nature.

To the best of our knowledge, optical properties of $BaPd_2As_2$ have not been studied yet, neither experimentally nor theoretically. A detailed theoretical study of various thermodynamic functions with pressure and temperature is still lacking. Study of optical properties yields valuable information which complements the electronic band structure calculations. In this paper we have presented a detailed investigation of optical functions and thermodynamical properties (bulk modulus, coefficient of volume expansion, heat capacity, and Debye temperature) as a function of temperature and pressure of $BaPd_2As_2$ for the first time. It should be mentioned that various ternary MAX phase compounds and other metallic ternaries with quasi-2D features exhibit superconductivity [16 – 18]. These layered nanolaminates often possess attractive optical properties suitable for applications. Hence, it is always instructive to study the optical parameters of layered ternary metallic systems keeping the possibility of potential applications in mind.

The structural and elastic properties are intimately related to the bonding characteristics of a crystal. This bonding and structural characteristics determines the features of phonon spectrum, and therefore set the energy scale for superconductivity. The electronic band structure and the electronic energy density of states at the Fermi energy are also quite important for emergence of superconductivity at low temperature. We have revisited these properties in this study. All the calculations were performed using the density functional theory (DFT) based *ab*-initio calculations.

Our study reveals conventional metallic characteristics for $BaPd_2As_2$. The electronic band structure calculations show several bands with varying degree of dispersive features crossing the Fermi level. The electronic density of states at the Fermi energy, $N(E_F)$, is derived primarily from the As $4p$ orbitals and Pd $4d$ orbitals with a small contribution from the Ba $5d$ electronic states. The elastic properties are highly anisotropic. The optical constants also show anisotropy with respect to the polarization of incident electromagnetic radiation.

The rest of this paper is organized as follows. Section 2 describes briefly the computational methodology. In Section 3, results of the computations are presented and analyzed. Important conclusions drawn from this investigation are discussed and summarized in Section 4.

## 2 Computational methodologies

The most popular practical approach to *ab*-initio modeling of structural and electronic properties of solids is the DFT with periodic boundary conditions. In this formalism the ground state of the crystalline system is found by solving the Kohn-Sham equation [19]. Experimental charge transport and heat capacity studies on $BaPd_2As_2$ [10, 15] showed clear metallic Fermi-liquid like behavior. Therefore, we have employed the generalized gradient approximation (GGA) for electron exchange correlation, using the CAmbridge Serial Total Energy Package (CASTEP) [20] designed to implement DFT based calculations. GGA relaxes the lattice constants due to the repulsive core-valence electron exchange correlation. Vanderbilt-type ultrasoft pseudopotentials were used to model the electron-ion interactions. This relaxes the norm-conserving criteria but produces a smooth and computation friendly pseudopotential. This saves computational time significantly without affecting the accuracy appreciably [21]. The functional form of the GGA has been chosen as the Perdew-Burke-Ernzerh of (PBE) type [22]. Density mixing is used to the electronic structure and Broyden Fletcher Goldfarb Shanno (BFGS) geometry optimization [23] has been employed to optimize the crystal structure for the given symmetry. The following electronic orbitals are used for Ba, Pd and As to derive the valence and the conduction bands, respectively: Ba $[5s^2 5p^6 6s^2]$, Pd $[4s^2 4p^6 4d^{10}]$ and As $[4s^2 4p^3]$. Periodic boundary conditions are used to determine the total energies of each cell. The trial wave functions are expanded in plane wave basis. The cut-off energy for the plane wave expansion is taken as 700 eV. *k*-point sampling within the Brillouin zone (BZ) for the compound under study has been carried out with $12 \times 12 \times 5$ special points in the Monkhorst-pack grid scheme [24]. This ensures the convergence of the energy – cell volume calculations. Geometry optimization is performed using the self-consistent convergence thresholds of $10^{-6}$ eV atom$^{-1}$ for the total energy, 0.03 eV Å$^{-1}$ for the maximum force, 0.05GPa for maximum stress, and $10^{-3}$ Å for maximum displacement. Elastic constants were calculated by the 'stress-strain' method as included within the CASTEP program. The bulk modulus, $B$ and the shear modulus, $G$ were obtained from the calculated single crystal elastic constants, $C_{ij}$. The electronic band structure features are calculated using the theoretically optimized geometry of BaPd$_2$As$_2$. All the optical constants are obtained by considering the electronic transition probabilities between different electronic orbitals. The imaginary part, $\varepsilon_2(\omega)$, of the complex dielectric function has been calculated within the momentum representation of matrix elements between occupied and unoccupied electronic states by employing the CASTEP supported formula expressed as

$$\varepsilon_2(\omega) = \frac{2e^2\pi}{\Omega\varepsilon_0} \sum_{k,v,c} \left| \left\langle \psi_k^c \left| \hat{u} \cdot \vec{r} \right| \psi_k^v \right\rangle \right|^2 \delta(E_k^c - E_k^v - E) \tag{1}$$

In the above expression, $\Omega$ is the volume of the unit cell, $\omega$ frequency of the incident electromagnetic wave (photon), $e$ is the electronic charge, $\psi_k^c$ and $\psi_k^v$ are the conduction and valence band wave functions at a given wave-vector $k$, respectively. The delta function ensures conservation of energy and momentum during the optical transition. The Kramers-Kronig transformations yield the real part $\varepsilon_1(\omega)$ of the dielectric function from the corresponding imaginary part $\varepsilon_2(\omega)$. Once these two parts are known, all the optical constants can be obtained from them [25]. This procedure has been followed by a great deal of earlier works to reliably calculate the frequency dependent optical parameters [26 – 28].

To study the pressure and temperature dependent thermodynamic properties, we have employed the energy-volume data calculated from the third-order Birch-Murnaghan equation of state [29] using the zero temperature and zero pressure equilibrium values of energy, volume, and bulk modulus obtained through the DFT calculations.

## 3 Results and analysis
### 3.1 Structural and elastic properties
As mentioned in the preceding section, the superconducting phase of BaPd$_2$As$_2$ crystallizes in the ThCr$_2$Si$_2$-type structure. The optimized crystal structure of is shown in Fig. 1. The optimization was done by minimizing the energy of the structure with respect to the volume of the unit cell. The structural ground state yields the fully relaxed lattice parameters of BaPd$_2$As. These theoretical lattice parameters are given in Table 1, together with the experimentally available values obtained from room temperature X-ray diffraction data [10] and those obtained from earlier *ab*-initio estimation [13]. The calculated lattice parameters show good agreement with the values found in earlier studies [10, 13]. The theoretical unit cell volume is slightly larger than the experimental one. This is not entirely unexpected since GGA slightly overestimates the lattice constants due to *softening* of the electronic orbitals. Local density approximation (LDA), on the other hand, underestimates the lattice parameters. The agreement with experimental values of the lattice parameters are in fact somewhat better for our estimates compared to those found by Shein et al. [13].

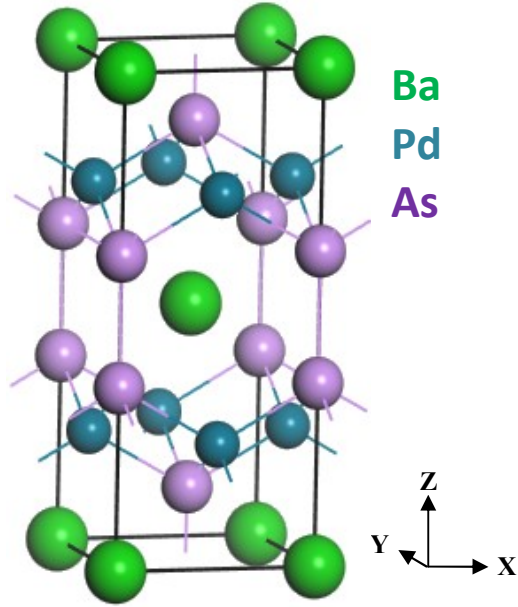

**Figure 1.** Crystal structure of $BaPd_2As_2$. The crystallographic directions are shown.

**Table 1.** Lattice parameters of $BaPd_2As_2$.

| $a$ (Å) | $c$ (Å) | $c/a$ | $V$ (Å$^3$) | Ref. |
|---------|---------|-------|-------------|------|
| 4.5013 | 10.4579 | 2.3233 | 211.8948 | This |
| 4.4890 | 10.3220 | 2.2993 | 207.9998* | [10]$^{experimental}$ |
| 4.5340 | 10.6562 | 2.3502 | 219.0611* | [13]$^{theoretical}$ |

*calculated from $V = a^2c$

$BaPd_2As_2$ has tetragonal structure. Therefore, following symmetry arguments, it has six independent single crystal elastic constants: $C_{11}$, $C_{12}$, $C_{13}$, $C_{33}$, $C_{44}$, and $C_{66}$. The calculated elastic constants are given in Table 2 with those estimated earlier [13] for comparison.

**Table 2.** Single crystal elastic constants, $C_{ij}$, of $BaPd_2As_2$ (all in GPa).

| $C_{11}$ | $C_{12}$ | $C_{13}$ | $C_{33}$ | $C_{44}$ | $C_{66}$ | Ref. |
|----------|----------|----------|----------|----------|----------|------|
| 150.6 | 52.8 | 65.2 | 114.5 | 50.7 | 30.2 | This |
| 129.0 | 35.4 | 46.5 | 67.1 | 41.3 | 23.6 | [13] |

It is seen from the above table, the calculated elastic constants are larger than those found by Shein et al. [13]. It is somewhat expected since one of the major factors determining the lattice parameters is the bonding strength among the atomic species. The larger theoretical values of the lattice parameters found by Shein et al. [13] (Table 1) imply weaker electronic bondings in the crystal. This should underestimate all the single crystal elastic constants and elastic moduli obtained theoretically.

Among the six independent elastic constants, $C_{11}$ and $C_{33}$ determine the response of the uniaxial strain. The other three are related to the shear dominated responses. In the present calculation, it is found that, $C_{11} > C_{33}$. This implies that the crystal is stiffer along the [100] and [010] directions in comparison to the stiffness along the [001] direction, as far as the uniaxial stresses are concerned. This is also directly linked to the strength of the bonding in the respective directions of the compound. Table 2 exhibits that $C_{44} > C_{66}$. The main implication of this finding is that, the [100] (010) shear should be easier than the [100] (001) shear for $BaPd_2As_2$. These qualitative and quantitative natures of the single crystal elastic constants reflect strongly the layered features of the crystal structure. Chemical bondings within the $ab$-plane are stronger than those along $c$-direction. One can investigate the state of mechanical stability of a given crystal structure by applying the well known Born-Huang [30] criteria. For tetragonal structure the stability criteria read:

$C_{11} - C_{12} > 0$; $C_{11} + C_{33} - 2C_{13} > 0$; $C_{ii} > 0$ (i = 1, 2, 3)

$2C_{11} + C_{33} + 2C_{12} + 4C_{13} > 0$ (2)

All these stability conditions are satisfied by the elastic constants of $BaPd_2As_2$.

The elastic moduli and the Poisson's ratio can be calculated from the values of $C_{ij}$. Table 3 shows the calculated elastic moduli and the Poisson's ratio of $BaPd_2As_2$. According to the Voigt approximation [31], isotropic bulk and shear moduli can be obtained from linear combinations of various elastic constants [31, 32 – 34]. We have denoted Voigt approximated bulk and shear moduli by $B_V$ and $G_V$, respectively. Using a different formalism, Reuss derived [35] different estimates for isotropic bulk and shear moduli from the single crystal elastic constants [32 – 34], denoted here by $B_R$ and $G_R$, respectively. Hill later proved that, the Voigt and Reuss approximated values are the upper and lower limits of the polycrystalline elastic moduli. A more realistic estimate of the bulk and shear moduli are therefore, the arithmetic averages given by, $B = (B_V + B_R)/2$ and $G = (G_V + G_R)/2$. Both Young's modulus, $E$, and Poisson's ratio, $n$, are related to the bulk modulus and to the shear modulus. From these relations [32 – 34] the values of these two parameters are obtained and presented in Table 3. The Pugh's ratio, defined as $B/G$, is an important mechanical indicator. We have also shown this ratio in the table given below.

**Table 3.** Elastic moduli (all in GPa), Pugh's ratio and Poisson's ratio of $BaPd_2As_2$.

| $B_V$ | $B_R$ | $B$ | $G_V$ | $G_R$ | $G$ | $B/G$ | $E$ | $n$ | Ref. |
|-------|-------|-----|-------|-------|-----|-------|-----|-----|------|
| 86.9 | 86.2 | 86.5 | 41.8 | 39.0 | 40.4 | 2.13 | 101.6 | 0.304 | This |
| 64.7 | 59.6 | 62.1 | 34.4 | 33.8 | 34.1 | 1.82 | 86.4 | 0.268 | [13] |

Compared to many other metallic and layered ternary compounds [16 – 18, 36, 37], the elastic moduli of $BaPd_2As_2$ is significantly lower, signifying its soft nature. It should be noted that the bulk modulus does not reflect the strength of a compound. Shear modulus is a better indicator. We have found that $G < B$. Therefore, the mechanical failure in $BaPd_2As_2$ should be controlled by the applied shear component. The Pugh's ratio is linked with the ductile/brittle behavior of a compound. The bulk modulus is a factor that measures the resistance to a volume change due to isotropic applied pressure and the shear modulus measures the resistance to plastic deformation. A high value of the Pugh's ratio is associated with ductility, whereas a low value corresponds to brittle nature. If $B/G > 1.75$, the material is expected to behave in a ductile manner; otherwise, the material is expected to be brittle. In addition, the ratio $B/G$ reflects the hardness of a material. The smaller the ratio $B/G$ is, the larger the hardness of the material. For $BaPd_2As_2$, we have found that $B/G > 1.75$. Therefore, this material should be ductile in nature. Poisson's ratio ($n$) plays an important part in assessing many mechanical properties of crystalline solids. It can predict the stability of solids against shear. Low value of $n$ is indicative of stability against shear [38]. Poisson's ratio is also related to the nature of interatomic forces in solids [39]. For solids where central force interaction dominates, $n$ resides in the ranges from 0.25 to 0.50 and for non-central force solids, $n$ lies outside of this. Furthermore, Poisson's ratio is an essential tool to predict the failure mode of crystalline solids [40, 41]. Those materials are expected to undergo brittle failure whose Poisson's ratio is less than a critical value of 0.26. With a Poisson's ratio greater than this critical value, a material undergoes ductile failure. Like Pugh's ratio prediction, the Poisson's ratio also predicts the brittle nature of $BaPd_2As_2$ compound. It also indicates that central force dominates in atomic bonding of $BaPd_2As_2$. In purely covalent crystals, the Poisson's ratio $\sim 0.10$, and in the completely metallic compounds, Poisson's ratio $\sim 0.33$. For the compound under study, $n = 0.304$. This implies that significant metallic bonding is present in $BaPd_2As_2$. The Young's modulus measures the resistance against expansive or compressive stress along the length. The value of $E$ is small (Table 3), showing that $BaPd_2As_2$ cannot withstand large tensile stress.

Study of elastic anisotropy is very useful for materials design, especially for compounds with layered structure. The elastic anisotropy has significant implications in engineering science due to its correlation with the possibility of creation and propagation of microcracks in the crystals. The calculated elastic anisotropy factors are listed in Table 4. The frequently used anisotropic factors for tetragonal crystal system are the shear anisotropic factors: $A_1$, $A_2$ and $A_3$. Moreover, the anisotropy indices for the bulk and shear moduli, $A_B$ and $A_G$, respectively are calculated. The universal anisotropic index, $A_U$, is also calculated. The following relations are used [42, 43]:

$A_1 = \frac{C_{44}(C_{11}+2C_{13}+C_{33})}{C_{11}C_{33}-C_{13}^2}$ for the (010) or (100) plane.

$A_2 = \frac{C_{44}(C_L+2C_{13}+C_{33})}{C_L C_{33}-C_{13}^2}$ for the (1$\bar{1}$0) plane, where $C_L = C_{66} + \frac{(C_{11}+C_{12})}{2}$.

$A_3 = \frac{2C_{66}}{C_{11}-C_{12}}$ for the (001) plane.

$A_B = \frac{B_V-B_R}{B_V+B_R} \times 100;\ A_G = \frac{G_V-G_R}{G_V+G_R} \times 100.$

$A_U = 5\frac{G_V}{G_R} + \frac{B_V}{B_R} - 6.$

Table 4 also includes the estimated isothermal compressibility, $\beta$, of BaPd$_2$As$_2$.

**Table 4.** Calculated indices of elastic anisotropy and compressibility (in GPa$^{-1}$) of BaPd$_2$As$_2$.

| $A_B$ | $A_G$ | $A^U$ | $A_1$ | $A_2$ | $A_3$ | $\beta$ | Ref. |
|-------|-------|-------|-------|-------|-------|---------|------|
| 0.4 | 3.5 | 0.37 | 1.5 | 1.8 | 0.62 | 0.0116 | This |

All the anisotropy indices depart significantly from unity. This indicates that the compound under investigation possesses highly anisotropic elastic property.

The difference between the two particular elastic constants, $(C_{12} - C_{44})$, defined as the Cauchy pressure [44], is an important physical parameter. This parameter provides us with an insight into charge density distribution and elastic response of solids [44]. Cauchy pressure can serve as an indicator of failure mode (ductility/brittleness) of crystalline solids. If Cauchy pressure is positive (negative), the material is expected to be ductile (brittle). The Cauchy pressure can also predict the nature of chemical bonding. Its positive value is indicative of metallic bonding and negative value corresponds to the directional covalent bonding with non-central angular character. In this study, we have found the Cauchy pressure for BaPd$_2$As$_2$ to be positive. Therefore, the compound is expected to behave as ductile with significant metallic bonding characteristic. This is consistent with the results obtained from the analysis of Pugh's and Poisson's ratios. Positive Cauchy pressure also implies that central force dominates in chemical bonding.

### 3.2 Electronic band structure of BaPd$_2$As$_2$

Electronic band structure calculations with the optimized crystal structure of BaPd$_2$As$_2$ are presented in this section. The electronic energy dispersion curves along the high-symmetry directions of the tetragonal Brillouin zone are shown in Fig. 2. The calculated total and partial density of states (TDOSs and PDOSs, respectively), as a function of energy, $(E - E_F)$, are presented in Figs. 3. The vertical straight line denotes the Fermi level, $E_F$, which has been set to zero. To understand the contribution of each atomic orbital to the TDOSs, we have estimated the PDOSs of Ba, Pd, and As atoms in $BaPd_2As_2$.

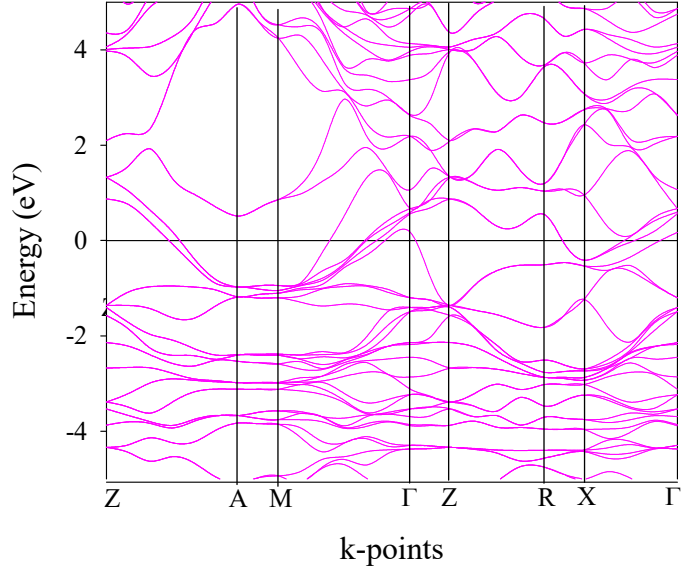

**Figure 2.** Electronic band structure of $BaPd_2As_2$ along the high symmetry directions of the *k*-space within the first Brillouin zone.

The band structure calculations reveal that a number of bands with varying degree of dispersion cross the Fermi level. This clearly illustrates the metallic nature of $BaPd_2As_2$. Most of the energy bands show electron-like features. The $E(k)$ curves crossing the Fermi level along the $Z - A$ and $M - \Gamma$ directions are highly dispersive. This indicates high mobility of the electrons in these directions. The $E(k)$ curve along $A - M$, on the other hand, is almost non-dispersive. This implies significant anisotropy in charge transport within and out of the *ab*-plane of $BaPd_2As_2$. The layered structure of this compound has led to notable anisotropy in the electronic band structure.

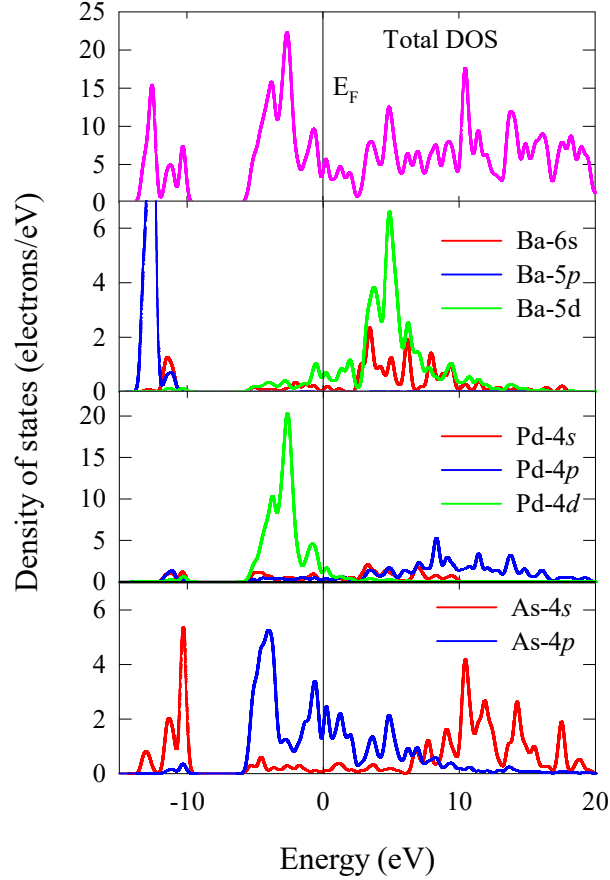

**Figure 3.** TDOS and PDOS of BaPd$_2$As$_2$. The vertical line shows the Fermi energy (set to zero).

To explore the contributions from different orbitals to the TDOS of BaPd$_2$As$_2$, we have calculated the atom-resolved partial electronic density of states. From Figs. 3, it is seen that the lowest energy band within -15 eV to ~ -10 eV is comprised of Ba-5$p$ and As-4$s$ electronic states. The contribution to the TDOS at the Fermi level comes primarily from the As-2$p$ orbital with some contributions from the Pd-4$d$ and Ba-5$d$ electronic orbitals. Therefore, hybridization among these three orbitals are expected to dominate charge transport and bonding properties of BaPd$_2$As$_2$. The higher energy band above the Fermi energy is formed by the Ba-5$d$, Ba-6$s$, Pd-4$p$, and As-4$s$ electronic states. The calculated TDOS at the Fermi level, $N(E_\text{F})$, is 4.2 states/eV-unit cell (containing two formula units). This value of $N(E_\text{F})$ agrees very well with those obtained by earlier studies [13].

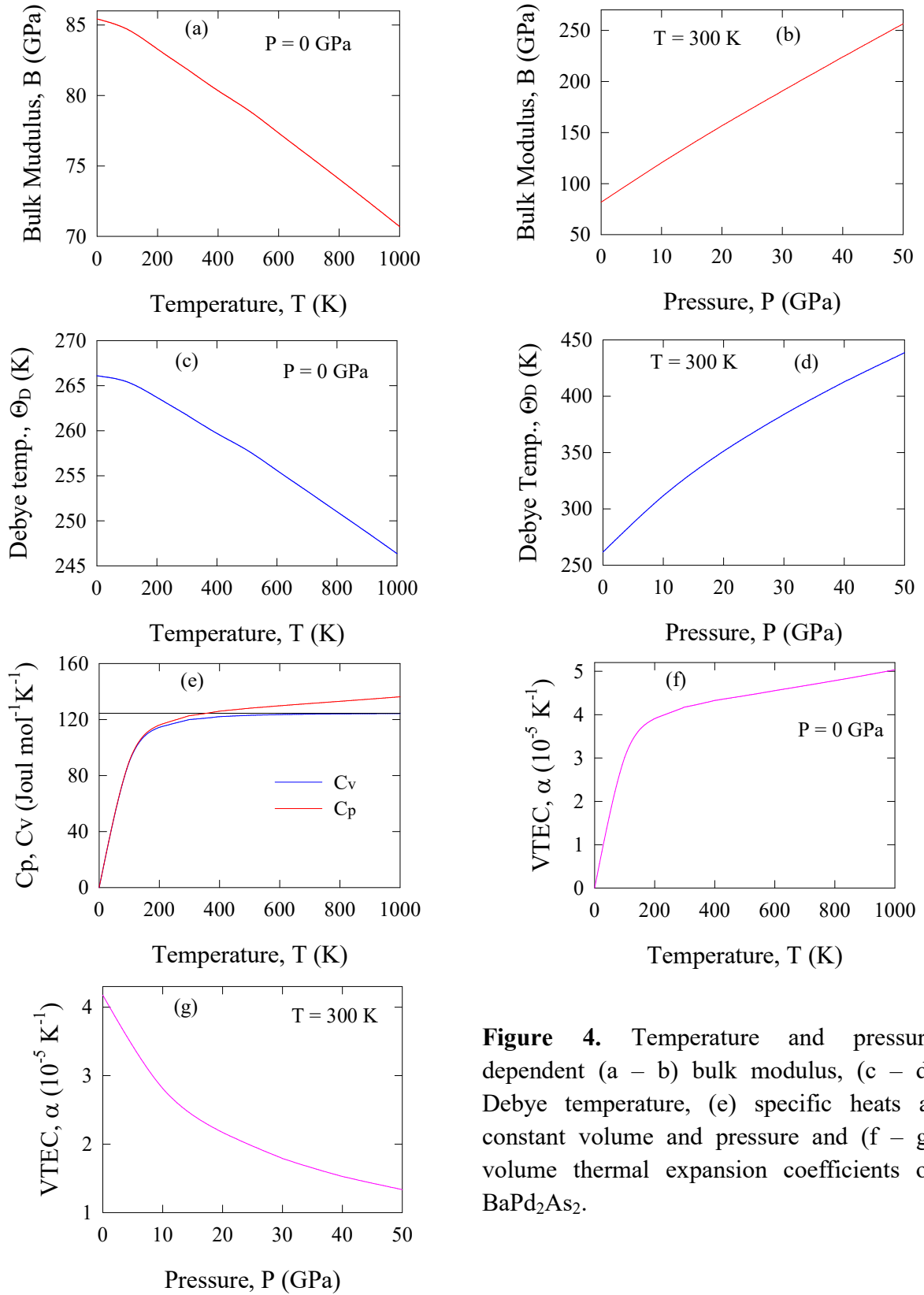

**Figure 4.** Temperature and pressure dependent (a – b) bulk modulus, (c – d) Debye temperature, (e) specific heats at constant volume and pressure and (f – g) volume thermal expansion coefficients of BaPd$_2$As$_2$.

Materials are often used at different pressures and temperatures. Therefore, study of the variation of different thermal parameters with pressure and temperature is important. Figs. 4a and 4b show the temperature and pressure dependences of bulk modulus of $BaPd_2As_2$. It is seen that at low temperature, the bulk modulus remains almost constant, a standard behavior related to the third law of thermodynamics. A sub-linear decrease of the bulk modulus at higher temperatures is observed (Fig. 4a). The bulk modulus increases quasi-linearly with pressure (Fig. 4b). This behavior is expected since external pressure tends to make the structure stiffer. Next, we have shown the $T$- and $P$-dependent behavior of the Debye temperature $\theta_D$, in Figs. 4c and 4d, respectively. Debye temperature is a fundamental material parameter. From the magnitude of the Debye temperature one can infer to a number of important physical parameters of solids namely, melting temperature, phonon specific heat, lattice thermal conductivity etc. It is also related to the bonding strength among the atoms present within the compound. In conventional superconductors, $\theta_D$ sets the characteristic energy scale for the phonons involved in Cooper pairing. From Figs. 4c and 4d, it follows that, initially the Debye temperature decreases very slowly with increasing temperature, a trend quite similar to that found for $T$-dependent bulk modulus. As the temperature increases further, $\theta_D$ starts to fall. This is due to two main effects – (i) as temperature increases number density of atomic oscillators decrease and (ii) the bonding strength among different atomic species also decreases. The pressure dependence of $\theta_D$ shows an opposite trend (Fig. 4d). The rate of increase of Debye temperature with pressure is high at low applied pressure; it decreases somewhat at higher temperatures. In this study, we have estimated $\theta_D$ under ambient condition using two different methodologies – via quasi-harmonic Debye approximation (QHDA) and using the following expressions which relate the sound velocity to the elastic moduli [36, 37].

$$\theta_D = \frac{h}{k_B}\left[\left(\frac{3n}{4\pi}\right)\frac{N_A \rho}{M}\right]^{1/3} v_m$$

where $h$ refers Planck's constant, $k_B$ is symbolized for Boltzmann's constant, $N_A$ is Avogadro's number, $\rho$ denotes mass density, $M$ is the molecular weight and $n$ is the number of atoms in the molecule. The average sound velocity $v_m$ in the polycrystalline material is determined by

$$v_m = \left[\frac{1}{3}\left(\frac{1}{v_l^3} + \frac{2}{v_t^3}\right)\right]^{-1/3}$$

where $v_l$ and $v_t$ are the longitudinal and transverse sound velocities in the solid and are obtained using the isotropic shear modulus $G$ and the bulk modulus $B$ as follows

$$v_l = \left[\frac{3B + 4G}{3\rho}\right]^{1/2}$$

and

$$v_t = \left[\frac{G}{\rho}\right]^{1/2}$$

The ambient condition value of $\theta_D$ was found to be 266 K and 269 K for QHDA and sound velocity method, respectively. Both our methods yield almost identical value $\theta_D$, giving an independent check of the accuracy.

The $T$-dependent behavior of the specific heat under constant volume, $C_V$, and constant pressure, $C_P$, are shown in Fig. 4e. The low-$T$ region of both $C_V$ and $C_P$ are characterized by the Debye $T^3$-law. At high temperatures $C_V$ approaches the classical Dulong-Petit limit. At high temperatures $C_P$ is markedly higher than $C_V$. This is somewhat expected in solids with high volume thermal expansion coefficient (VTEC). The $T$- and $P$-dependent VTEC of $BaPd_2As_2$ are shown in Figs. 4f and 4g, respectively.

### 3.4 Optical properties of $BaPd_2As_2$

Optical properties of a material determine response of the compound to the incident electromagnetic radiation. The response to visible spectra is particularly important from the view of optoelectronic device applications. This response to the incident electromagnetic wave is completely determined by the various energy dependent (frequency) optical parameters, namely the real and the imaginary parts of the dielectric constants, $\varepsilon_1(\omega)$ and $\varepsilon_2(\omega)$, respectively, real part of refractive index $n(\omega)$, extinction coefficient $k(\omega)$, real and imaginary parts of the optical conductivity $\sigma_1(\omega)$ and $\sigma_2(\omega)$, respectively, reflectivity $R(\omega)$, and the absorption coefficient $\alpha(\omega)$. The estimated optical parameters for $BaPd_2As_2$ for incident photon energies up to 20 eV, with electric field polarization vectors along [100] and [001] directions, are presented in Figs. 5. The polarization dependent response of the optical constants yields valuable information regarding optical and electronic anisotropy.

Fig. 5a illustrates the real and imaginary parts of the dielectric constants. Both $\varepsilon_1(\omega)$ and $\varepsilon_2(\omega)$ exhibit metallic characteristics. For metallic compounds, in the low frequency region where $\omega\tau << 1$, $\varepsilon_2(\omega)$ dominates the optical behavior since $\varepsilon_2(\omega) \sim \sigma_1(\omega)/\omega$. $\sigma_1(\omega)$ is finite at low frequencies in metallic compounds. On the other hand, when $\omega\tau >> 1$ (high frequency region), the real part approaches unity and the imaginary part becomes very small. This region is characterized by the inductive nature of electronic motion induced by high-frequency incident electromagnetic radiation. Here, $\tau$ denotes the electronic relaxation time. It is seen that the real part of the dielectric constant crosses zero from below at $\sim$ 15 eV and approaches unity and the imaginary part falls sharply and flattens to a low value at the same energy. Therefore, the material is expected to become transparent to incident electromagnetic radiation above 15 eV, the plasma frequency, $\omega_P$. The structures seen in $\varepsilon_1(\omega)$ and $\varepsilon_2(\omega)$ are due to electronic transitions between states. For example the broad peak around 5 eV arises due to electronic transition between the Pd-4$d$, As-4$p$ orbitals near the Fermi energy and the Ba-4$d$ electronic state above the Fermi level.

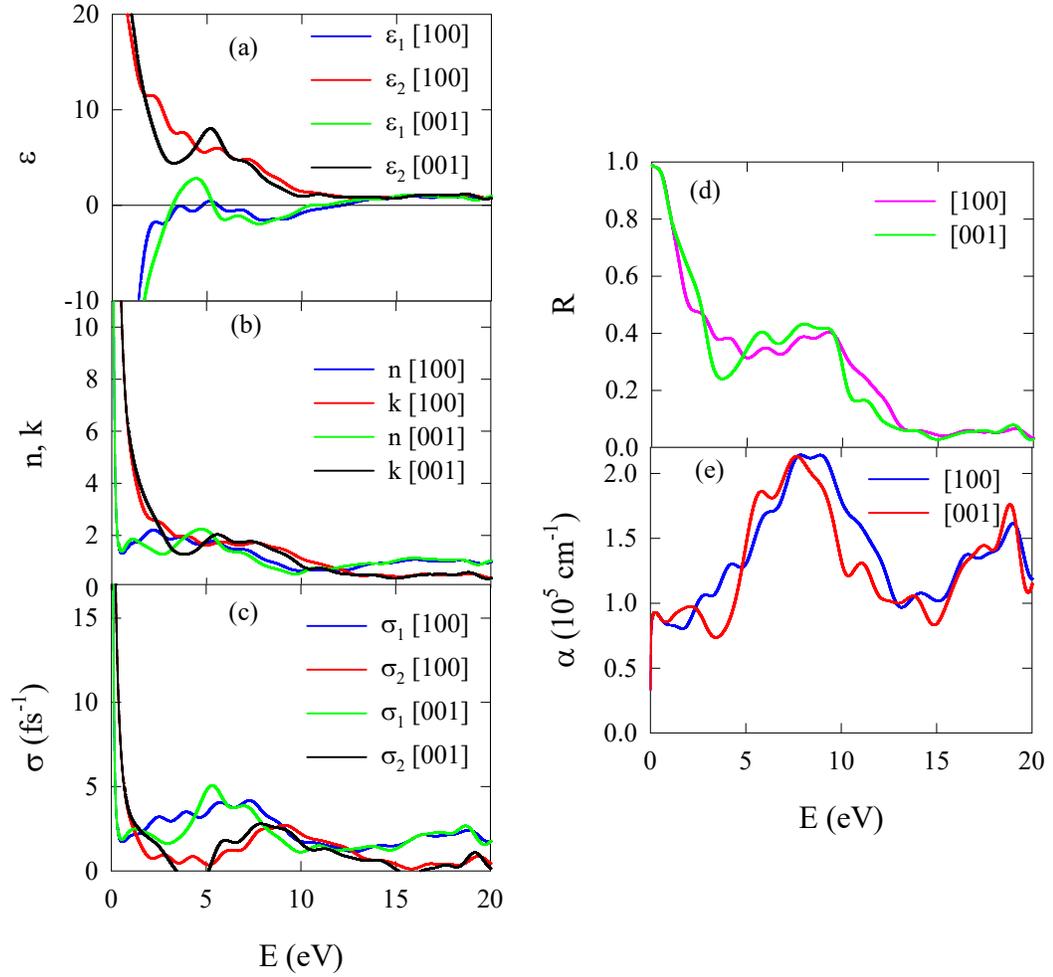

**Figure 5.** The frequency dependent (a) dielectric constant, (b) refractive index, (c) optical conductivity, (d) reflectivity, and (e) absorption coefficient of BaPd$_2$As$_2$ with electric field polarization vectors along [100] and [001] directions.

Refractive index is of critical importance for photonic applications, such as for optical wave-guides. The real and the imaginary parts of the refractive index, as a function of energy, are shown in Fig. 5b. The real part of the refractive index determines the phase velocity of the electromagnetic wave inside the compound, while the imaginary part, often termed as extinction coefficient, gives the measure of the amount attenuation when the electromagnetic wave travels through the material. The optically dispersive character of a material is completely determined by the frequency dependent refractive index. Since both the real and imaginary part of the refractive index are related to the dielectric constant via the relations $\varepsilon_1(\omega) = n^2(\omega) - k^2(\omega)$ and $\varepsilon_2(\omega) = 2n(\omega)k(\omega)$, at high energies both real and imaginary parts of the refractive index becomes small as exhibited in Fig. 5b.

The real and imaginary part of the calculated optical conductivity is shown in Fig. 5c. Optical conductivities are finite at zero frequency. This is a hall mark of metallic conductivity consistent with the band structure calculations. The optical conductivities decrease with increasing frequency – a common feature of metallic systems.

The reflectivity profile is shown in Fig. 5d. Reflectivity is high in the infrared and near visible region. $R(\omega)$ decreases sharply with energy and flattens in the ultraviolet region over an extended energy range (5 – 10 eV). In this energy region the reflectivity remains ~ 40%. Further increase in the frequency of the incident electromagnetic wave results in a steady decrease of the reflectivity. $R(\omega)$ becomes almost zero for energies greater than ~ 15 eV - the plasma edge.

The absorption spectrum is presented in Fig. 5e. The peaks in this spectrum follow from interband transitions. The monotonic features are due to intraband optical transitions. Finite absorption at low energies is due to free carriers which show once again the metallic character of $BaPd_2As_2$. It is interesting to note that this compound absorbs ultraviolet radiation quite effectively.

All the calculated optical parameters show significant dependence of the state of electric field polarization. The material under study, therefore, possesses noticeable optical anisotropy.

## 4 Discussion and conclusions

First-principles DFT based calculations have been used to investigate the elastic, electronic, thermodynamic, and optical properties layered ternary $BaPd_2As_2$ superconductor with $ThCr_2Si_2$-type structure. The pressure and temperature dependent thermodynamic parameters have been studied using the QHDA for the first time. The energy dependent optical parameters have been calculated with different electric field orientations, also for the first time. The optimized lattice parameters agree quite well with earlier studies [10, 13]. The elastic constants and the moduli reveals that this compound is relatively soft compared to many other layered ternaries [16 – 18, 36, 37]. The bulk modulus $B$ together with elastic constant $C_{44}$ can assess the machinability of a compound with the machinability index, $\mu_M = B/C_{44}$, as defined by Sun et al. [45]. The calculated $\mu_M$ of $BaPd_2As_2$ is 1.71, which implies that this material is fairly machinable. $BaPd_2As_2$ possesses significant elastic anisotropy. The Pugh's ratio, Poisson's ratio, Cauchy pressure, all these indicators imply ductile nature of $BaPd_2As_2$ with significant metallic bonding characteristics. The Cauchy pressure indicates that central force dominates in atomic bondings. Debye temperature extracted from the QHDA and elastic constants are almost identical. It should be noted that both Kudo et al. [11] and Abdel-Hafiez et al. [15] have obtained $\theta_D$ from the analysis of experimental heat capacity data which turns out to be ~ 144 K. This value is significantly lower than the value obtained in this study. Chen et al. [14], on the other hand, has calculated the Debye temperature from the analysis of phonon dispersion and found a value of 213 K. Johnston et al. [46] have found $\theta_D = 273$ K from low-$T$ fit to the heat capacity data. A wide variation among the reported values is a matter of concern and requires further attention.

For a given value of electron-phonon coupling constant, $\lambda_{e\text{-ph}}$, the superconducting transition temperature is directly proportional to the Debye temperature. The electron phonon coupling constant can be expressed as $\lambda_{\text{e-ph}} = N(E_F)V_{\text{e-ph}}$, where $V_{\text{e-ph}}$ is the matrix element of electron-phonon interaction responsible for Cooper pairing. We have estimated $\lambda_{\text{e-ph}}$ for BaPd$_2$As$_2$ using the experimental $T_c$ value via McMillan $T_c$-equation [47]. The value of Coulomb pseudopotential, $\mu^*$ has been taken as 0.13, a typical value for superconducting compounds [48] and $\theta_D = 266$ K. The calculated value turns out to be, $\lambda_{\text{e-ph}} = 0.615$. This makes BaPd$_2$As$_2$ a moderately coupled electron-phonon superconductor. The matrix element of electron-phonon interaction should be relatively high since the density of states at the Fermi level is low. It should be noted that Chen et al. [14] have theoretically calculated the values of $\lambda_{\text{e-ph}}$ for $A$Pd$_2$As$_2$ [$A$ = Ba, Sr, Ca] ternaries which lie within 0.60 to 1.0. Abdel-Hafiez et al. [15] have also theoretically estimated $\lambda_{\text{e-ph}}$ for BaPd$_2$As$_2$ under ambient condition which turns out to be 0.70, a value quite close to our estimation. It is worth pointing out that the Fermi level of BaPd$_2$As$_2$ lies within a dip in the electronic density of states profile. Slight shift of the Fermi level to lower energy can increase $N(E_F)$ significantly. This can be accomplished by suitable atomic substitution and increase the $T_c$ markedly.

Band structure calculations reveal clear metallic behavior. The bands within the *ab*-plane are dispersive in nature, while the bands along the *c*-direction are less dispersive. This implies that the structurally layered feature has notable effect on charge transport within and out of the layer directions. The optical properties are also anisotropic. The state of polarization of the electric field has appreciable effect on the magnitude of all the optical constants. The reflectivity is quite high in the infrared and near-visible region. It becomes non-selective in the energy range 5 – 10 eV in the ultraviolet region. The absorption coefficient is high and non-selective in the ultraviolet region. All these features can be useful for optoelectronic device applications.

**List of references**

1. Y. Kamihara, T. Watanabe, M. Hirano, H. Hosono, Iron-based layered superconductor La[O$_{1-x}$F$_x$]FeAs (x = 0.05-0.12) with $T_c$ = 26 K, J. Am. Chem. Soc. 130 (2008) 3296-3297.

2. X. H. Chen, T. Wu, G. Wu, R. H. Liu, H. Chen, D. F. Fang, Superconductivity at 43 K in SmFeAsO$_{1-x}$F$_x$, Nature 453 (2008) 761.

3. J. Paglione, R. L. Green, High-temperature superconductivity in iron-based materials, Nat. Phys. 6 (2010) 645-658.

4. P. J. Hirschfeld, M. M. Korshunov, I. I. Mazin, Gap symmetry and structure of Fe-based superconductors, Rep. Prog. Phys. 74 (2011) 124508.

5. Y. Bang, G. R. Stewart, Superconducting properties of the $s^{\pm}$-wave state: Fe-based superconductors, J. Phys.:Condens. Matter 29 (2017) 123003.

6. Z. D. Leong, P. Phillips, The effects of Coulomb interactions on the superconducting gaps in iron-based superconductors, Phys. Rev. B 93 (2016)155159.

7. A. V. Chubukov, P. J. Hirschfeld, Iron-based superconductors, seven years later, Phys. Today 68(6) (2015) 46.

8. R.M. Fernandes, A.V. Chubukov, J. Schmalian, What drives nematic order in iron-based superconductors?, Nat. Phys. 10 (2014) 97.

9. D. C. Johnston, The puzzle of high temperature superconductivity in layered iron pnictides and chalcogenides, Adv. Phys. 59 (2010) 803.

10. Q. Guo, J. Yu, B. Ruan, D. Chen, X. Wang, Q. Mu, B. Pan, G. Chen, Z. Ren, Superconductivity at 3.85 K in $BaPd_2As_2$ with the $ThCr_2Si_2$-type structure, EPL 113 (2016) 17002.

11. K. Kudo, Y. Yamada, T. Takeuchi, T. Kimura, S. Ioka, G. Matsuo, Y. Kitahama, M. Nohara, Strong-coupling superconductivity in $BaPd_2As_2$ induced by soft phonons in the $ThCr_2Si_2$-type polymorph, J. Phys. Soc. Jpn. 86 (2017) 063704.

12. V. K. Anand, H. Kim, M. A. Tanatar, R. Prozorov, Superconducting and normal-state properties of $APd_2As_2$ (A = Ca, Sr, Ba) single crystals, Phys. Rev. B 87 (2013) 224510.

13. I. R. Shein, S. L. Skornyakov, V. I. Anisimov, A. L. Ivanovskii, Elastic and Electronic Properties of Superconducting $CaPd_2As_2$ and $SrPd_2As_2$ vs. Non-superconducting $BaPd_2As_2$, J. Supercond. Nov. Magn. 27(1) (2014) 155-161.

14. J. Chen, Phonons and Electron–Phonon Coupling of Newly Discovered $ThCr_2Si_2$-Type Superconductor $BaPd_2As_2$: A Comparison Study with $Sr(Ca)Pd_2As_2$, J. Supercond. Nov. Magn. 29 (2016) 1219-1225.

15. M. Abdel-Hafiez, Y. Zhao, Z. Huang, C.-w. Cho, C. H. Wong, A. Hassen, M. Ohkuma, Y.-W. Fang, B.-J. Pan, Z.-A. Ren, A. Sadakov, A. Usoltsev, V. Pudalov, M. Mito, R. Lortz, C. Krellner, W. Yang, High-pressure effects on isotropic superconductivity in the iron-free layered pnictide superconductor $BaPd_2As_2$, Phys. Rev. B 97 (2018) 134508.

16. M. A. Hadi, M. Roknuzzaman, F. Parvin, S. H. Naqib, A. K. M. A. Islam, M. Aftabuzzaman, New MAX phase superconductor $Ti_2GeC$: a first-principles study, Journal of Scientific Research 16(1) (2014) 11.

17. M. A. Hadi, M. A. Alam, M. Roknuzzaman, M. T. Nasir, A. K. M. A. Islam, S. H. Naqib, Structural, elastic, and electronic properties of recently discovered ternary silicide superconductor $Li_2IrSi_3$: An *ab-initio* study, Chinese Physics B 24(11) (2015) 117401.

18. M. A. Hadi, M. T. Nasir , M. Roknuzzaman, M. A. Rayhan, S. H. Naqib, A. K. M. A. Islam, First-principles prediction of mechanical and bonding characteristics of new $T_2$ superconductor $Ta_5GeB_2$, Physica Status Solidi (b) 253(10) 2020 (2016).

19. W. Kohn, L. J. Sham, Self-consistent equations including exchange and correlation effects, Phys. Rev. 140 (1965) A1133.

20. S. J. Clark, M. D. Segall, C. J. Pickard, P. J. Hasnip, M. I. J. Probert, K. Refson, M. C. Payne, First principles methods using CASTEP, Z. Kristallographie 220 (2005) 567-570.

21. P. Barua, M. M. Hossain, M. A. Ali, M. M. Uddin, S. H. Naqib, A. K. M. A. Islam, Effects of transition metals on physical properties of $M_2BC$ (M = V, Nb, Mo and Ta): A DFT calculation, Journal of Alloys and Compounds 770 (2019) 523.

22. J. P. Perdew, K. Burke, M. Ernzerhof, Generalized gradient approximation made simple, Phys. Rev. Lett. 77 (1996) 3865.

23. T. H. Fischer, J. Almlof, General methods for geometry and wave function optimization, J. Phys. Chem. 96 (1992) 9768.

24. H. J. Monkhorst, J. D. Pack, Special points for Brillouin-zone integrations, Phys. Rev. B 13 (1976) 5188.

25. N. W. Ashcroft, N. D. Mermin, Solid State Physics, Holt-Saunders (international edition), USA, 1976.

26. A. Chowdhury, M. A. Ali, M. M. Hossain, M. M. Uddin, S. H. Naqib, A. K. M. A. Islam, Predicted MAX phase $Sc_2InC$: Dynamical stability, vibrational and optical properties, Physica Status Solidi (b) 255(3) (2018)1700235.

27. M. A. Ali, M. T. Nasir, M. R. Khatun, A. K. M. A. Islam, S. H. Naqib, An *ab initio* investigation of vibrational, thermodynamic, and optical properties of $Sc_2AlC$ MAX compound, Chinese Physics B 25(10) (2016) 103102.

28. M. Roknuzzaman, M. A. Hadi, M. J. Abden, M. T. Nasir, A. K. M. A. Islam, M. S. Ali, K. Ostrikov, S. H. Naqib, Physical properties of predicted $Ti_2CdN$ versus existing $Ti_2CdC$ MAX phase: An *ab initio* study, Computational Materials Science 113 (2016) 148.

29. F. Birch, Finite strain isotherm and velocities for single-crystal and polycrystalline NaCl at high pressures and 300 K J. Geophys. Res.: Solid Earth 83 (1978) 1257-1268.

30. M. Born, K. Huang, Dynamical Theory of Crystal Lattices, Oxford University Press, UK, 1998.

31. W. Voigt, Lehrbuch der Kristallphysik, Teubner, Leipzig, 1928.

32. M. A. Hadi, S. H. Naqib, S. R. G. Christopoulos, A. Chroneos, A. K. M. A. Islam, Mechanical behavior, bonding nature and defect processes of $Mo_2ScAlC_2$: A new ordered MAX phase, Journal of Alloys and Compounds 724 (2017) 1167.

33. M. A. Ali, M. M. Hossain, M. A. Hossain, M. T. Nasir, M. M. Uddin, M. Z. Hasan, A. K. M. A. Islam, S. H. Naqib, Recently synthesized $(Zr_{1-x}Ti_x)_2AlC$ $(0 \leq x \leq 1)$ solid solutions: Theoretical study of the effects of M mixing on physical properties, Journal of Alloys and Compounds 743 (2018) 146.

34. M. A. Hadi, M. Roknuzzaman, A. Chroneos, S. H. Naqib, A. K. M. A. Islam, R. V. Vovk, K. Ostrikov, Elastic and thermodynamic properties of new $(Zr_{3-x}Ti_x)AlC_2$ MAX-phase solid solutions, Computational Materials Science 137 (2017) 318.

35. A. Reuss, Berechnung der Fließgrenze von Mischkristallen auf Grund der Plastizitäts bedingung für Einkristalle, Z. Angew. Math. Mech. 9 (1929) 49-58.

36. F. Parvin, S. H. Naqib, Elastic, thermodynamic, electronic, and optical properties of recently discovered superconducting transition metal boride NbRuB: An *ab-initio* investigation, Chinese Physics B 26(10) (2017) 106201.

37. M. A. Ali, M. A. Hadi, M. M. Hossain, S. H. Naqib, A. K. M. A. Islam, Theoretical investigation of structural, elastic, and electronic properties of ternary boride MoAlB, Physica Status Solidi(b) 254 (2017) 1700010.

38. P. Ravindran, L. Fast, P.A. Korzhavyi, B. Johansson, J. Wills, O. Eriksson, Density functional theory for calculation of elastic properties of orthorhombic crystals: Application to $TiSi_2$, J. Appl. Phys. 84 (1998) 4891.

39. O. L. Anderson and H. H. Demarest Jr., Elastic constants of the central force model for cubic structures: Polycrystalline aggregates and instabilities, J. Geophys. Res. 76(1971) 1349.

40. I. N. Frantsevich, F. F. Voronov, S. A. Bokuta, Elastic Constants and Elastic Moduli of Metals and Insulators Handbook (Naukova Dumka, Kiev, 1983), pp. 60–180.

41. G. Vaitheeswaran, V. Kanchana, A. Svane, and A. Delin, Elastic properties of $MgCNi_3$ - a superconducting perovskite, J. Phys.: Condens. Matter 19(2007) 326214.

42. M-X. Zeng, R-N. Wang, B-Y. Tang, L-M. Peng, W-J. Ding, Elastic and electronic properties of t126-type $Mg_{12}RE$ (RE = Ce, Pr and Nd) phases, Modelling Simul. Mater. Sci. Eng. 20 (2012) 035018.

43. Z-J. Liu, S-Q. Duan, J. Yan, X-W. Sun, C-R. Zhang,Y-D. Chu, Theoretical investigations of the physical properties of tetragonal $CaSiO_3$ perovskite, Solid State Commun. 150 (2010) 943.

44. M. E. Eberhart, T. E. Jones, Cauchy pressure and the generalized bonding model for nonmagnetic bcc transition metals, Phys. Rev. B 86 (2012)134106  and references therein.

45. Z. Sun, D. Music, R. Ahuja, J. M. Schneider, Theoretical investigation of the bonding and elastic properties of nanolayered ternary nitrides, Phys. Rev. B 71 (2005) 193402.

46. V. K. Anand, H. Kim, M. A. Tanatar, R. Prozorov, D. C. Johnston, Superconducting and normal-state properties of $APd_2As_2$ (A = Ca, Sr, Ba) single crystals, Phys. Rev. B 87 (2013) 224510.

47. W. L. McMillan, Transition Temperature of Strong-Coupled Superconductors, Phys. Rev. 167 (1968) 331.

48. A. K. M. A. Islam, S. H. Naqib, Possible explanation of high-$T_c$ in some 2D cuprate superconductors, J. Phys. Chem. Solids 58 (1997) 1153.